# A New Homogenization-Free Boundary Condition Towards Aperiodic Metasurface Design Using Full-Wave Surrogate Models of Printed Circuits

**J. Budhu[1], R. Pestourie[2]**

[1] Bradley Department of Electrical and Computer Engineering, Virginia Tech, Blacksburg, VA, USA
[2] School of Computational Science and Engineering, Georgia Tech, Atlanta, GA, USA
jbudhu@vt.edu, rpestourie3@gatech.edu

*Abstract* – A new homogenization-free boundary condition is introduced for the design of metasurfaces. The boundary condition, linking the tangential electric field to the induced surface current density within a unit cell, is described as a matrix equation containing a surrogate model of a printed circuit. Full-wave simulations are performed to construct the data needed to capture the physics of a printed circuit's response to exciting fields in terms of its induced current density. The matrix equation takes the form of a boundary condition which can be included in design and optimization algorithms incorporating mutual coupling in the form of an impedance matrix. Modelling the metasurface in this manner allows for direct design of the realizable model avoiding any homogenization or locally periodic approximations leading to a paradigm shift in metasurface design approaches. The surrogate model's construction is described and its use in a 50 printed circuit example is provided.

## I. INTRODUCTION

Traditionally, metasurfaces are modeled using the Generalized Sheet Transition Condition or GSTC [1]. For metasurfaces invariant in one dimension which support only electric currents driven by electric fields, the GSTC reduces to the scalar impedance boundary condition $E_{tan}^{tot} = \eta_s J_s$ [1]. This boundary condition relies on the homogenization of the metasurfaces constituent meta-atoms. Homogenization refers to averaging the response of a unit cell (which in general contains a complex geometry) and assigning simple proportionality constants (impedances, admittances, and coupling coefficients) between induced currents and averaged fields (in the scalar case above, $\eta_s$). This way the metasurface can be described, analyzed, and designed in a reduced dimensional space. After a design is completed in the reduced dimensional space, the proportionality constants must be translated back to complex geometries by way of an extraction method to realize the metasurface. It isn't always the case that the homogenized metasurface is realizable as an array of meta-atoms which duplicates the performance of the homogenized design. For example, in [2], conducting baffles had to be introduced to recover the performance of the homogenized metasurface. Or in [3], an additional full wave optimization had to be performed on the realized metasurface to bring the performance back in agreement with the homogenized design. This isn't always feasible for larger metasurfaces. One approach which attempts to solve the duplication problem is described in [4] for metasurfaces modeled using matrix equations [3,5]. However, this approach is costly as each unit cell must be designed one-at-a-time using full-wave simulation-based optimizations. Furthermore, the optimizations themselves may not always converge. Rather than attempting to recreate the performance of the homogenized metasurface using printed circuits *post-design*, here we present an approach to incorporate the printed circuits response *intra-design* using full-wave data driven surrogate models. Each surrogate model describes the underlying physics of a printed circuit and hence in theory can substitute in for the actual printed circuit. The surrogate models, derived from full-wave simulations of printed circuits, are cast as a new boundary condition which replaces the GSTC and hence avoids the need for homogenization all together. Thus, the new design approach can be described as homogenization-free. Similar to the GSTC which can be written as a diagonal matrix equation (when $k$ samples are placed along a unit cell containing a homogenized impedance sheet [3])

$$\left[ E_{\tan}^{tot} \right]_{k \times 1} = \left[ diag(\eta_s) \right]_{k \times k} \left[ J_s \right]_{k \times 1} \quad (1)$$

the surrogate model-based boundary condition links the total tangential electric field to the induced surface current density along a printed circuits surface

$$\left[ E_{\tan}^{tot} \right]_{k \times 1} = \left[ A(l) \right]_{k \times k} \left[ J_s \right]_{k \times 1} \quad (2)$$

where $[A(l)]$ is a $k \times k$ matrix learned from full-wave simulations. Its elements are a function of $l$, the geometrical variable parameterizing the printed circuit (see Fig. 1a). For each $l$ a different matrix $A$ results. The learning of $[A(l)]$ need only be done once and can be integrated into a general metasurface design algorithm thereafter. The advantage of the surrogate model is its inherent order reduction capturing the physics of the printed circuit in a finite number of parameters, in this case $k$ of them. In general, $k \ll C$, where $C$ indicates the degrees of freedom required for a full wave simulation of the printed circuit itself. This reduction of order leads to an evaluation of the induced current density on a printed circuit given the exciting field of 5 orders of magnitude faster than the full-wave simulations used to generate the surrogate model.

The new boundary condition (2) enters into the matrix equation in the same way as the homogenization-based GSTC of (1) does [3,5]. For example, a metasurface containing 3 interdigitated capacitor (IDC) printed circuits would be modeled using the matrix equation obtained from the moment method of [3-5]

$$\begin{bmatrix} [V_1]_{k\times 1} \\ [V_2]_{k\times 1} \\ [V_3]_{k\times 1} \end{bmatrix} = \begin{bmatrix} [A(l_1)]_{k\times k} & [0]_{k\times k} & [0]_{k\times k} \\ [0]_{k\times k} & [A(l_2)]_{k\times k} & [0]_{k\times k} \\ [0]_{k\times k} & [0]_{k\times k} & [A(l_3)]_{k\times k} \end{bmatrix} \begin{bmatrix} [I_1]_{k\times 1} \\ [I_2]_{k\times 1} \\ [I_3]_{k\times 1} \end{bmatrix} + \begin{bmatrix} [Z_{11}]_{k\times k} & [Z_{12}]_{k\times k} & [Z_{13}]_{k\times k} \\ [Z_{21}]_{k\times k} & [Z_{22}]_{k\times k} & [Z_{23}]_{k\times k} \\ [Z_{31}]_{k\times k} & [Z_{32}]_{k\times k} & [Z_{33}]_{k\times k} \end{bmatrix} \begin{bmatrix} [I_1]_{k\times 1} \\ [I_2]_{k\times 1} \\ [I_3]_{k\times 1} \end{bmatrix} \quad (3)$$

where $[V_p]$ represents the incident field on element $p$, $[Z_{pq}]$ represents the impedance matrix coupling element $p$ to $q$ calculated using Green's functions, and $[I_q]$ represents the current density on element $q$. The matrix equation (3) can either be solved directly given a metasurfaces geometry (the $l's$ of each meta-atom's printed circuit) or can be integrated into an optimization loop where each meta-atoms $l$ acts as an optimization variable. However now, the design and optimization is directly done on the realized model rather than a homogenized approximation. Thus, the traditionally required non-guaranteed final translation step to a realizable metasurface is completely avoided. Instead, when the optimizer converges, the physical design is the output. This is also in contrast to other approaches which utilize surrogate models and matrix equations where the design is still done on the homogenized model and only post-design is the physical structure created [6]. We describe the development of the Surrogate model next, followed by an example of a metasurface simulation utilizing an array of coupled surrogate models as in (3).

## II. Development of the Surrogate Model

To develop the surrogate model, COMSOL Multiphysics was used to impress 8 different choices of an inhomogeneous background electric field to a medium containing a single 8-finger IDC suspended in free space. The IDC is $(2/3)\lambda_0/10$ wide and $1.5\lambda_0/10$ tall (see Fig. 1a). For each of the 8 background fields, a total of 96 different $l$'s ranging from 0.0002mm to 0.004mm were simulated. As metasurfaces invariant in one dimension were targeted, the IDC is placed within a parallel plate waveguide and hence the plate separation is $1.5\lambda_0/10$. Both the induced surface current density on the surface of the IDC and the total electric field within its gap are exported and averaged vertically then sampled horizontally at $k = 8$ locations corresponding to the 8 fingers of the IDC (see (2) of [4]). This provides the vectors $[J_s]_{k\times 1}$ and $[E_{tan}^{tot}]_{k\times 1}$ of (2). The matrix $[A(l)]_{k\times k}$ is then inferred from this data via a least square fit. The physically meaningful $[J_s]_{k\times 1}$ and $[E_{tan}^{tot}]_{k\times 1}$, that are computed indirectly from the choice of background electric field, showed multicollinearity. However, a ridge regularization makes the fitting problem well posed. The closed form formula for the optimal matrix is

$$\forall l, A(l) = \sum_{i=1}^{8} E_i(l) J_i(l)^T \left( \sum_{i=1}^{8} J_i(l) J_i(l)^T - \alpha I \right)^{-1} \quad (4)$$

Where $\alpha = 0.0005$ is the coefficient of the ridge regularization, $T$ denotes transpose operator, and $I$ is the identity matrix. Note that (4) is different from the usual normal equation because we fit the matrix instead of the vector. For the design and optimization of the metasurface, each element of the matrix is interpolated as a function of the meta-atom parameter $l$ for the fast evaluation of the function and its derivative. Using this surrogate model not only results in a $10^5$ computational speedup, but also enables to evaluate $[E_{tan}^{tot}]_{k\times 1}$ from $[J_s]_{k\times 1}$, which COMSOL can't simulate directly. The reduced order model of the matrix fit solves the curse of dimensionality by removing the $k$ physical degrees of freedom $[J_s]_{k\times 1}$ from the surrogate model parameters. This leaves a surrogate model which is only a function of a single geometric parameter $l$. $[E_{tan}^{tot}]_{k\times 1}$ is evaluated by interpolating the $k^2$ matrix coefficients along the dimension of the meta-atom parameter $l$. A single surrogate model with $k + n$ inputs, where $n = 1$ is the number of geometric parameters, would require training data exponential in $(k + n)$ [7]. Here, we train $k^2$ surrogate models with $n$ inputs, which only need training data that is exponential in $n$. This surrogate model informed by the reduced order model is thus very efficient with both data and computational resources [8].

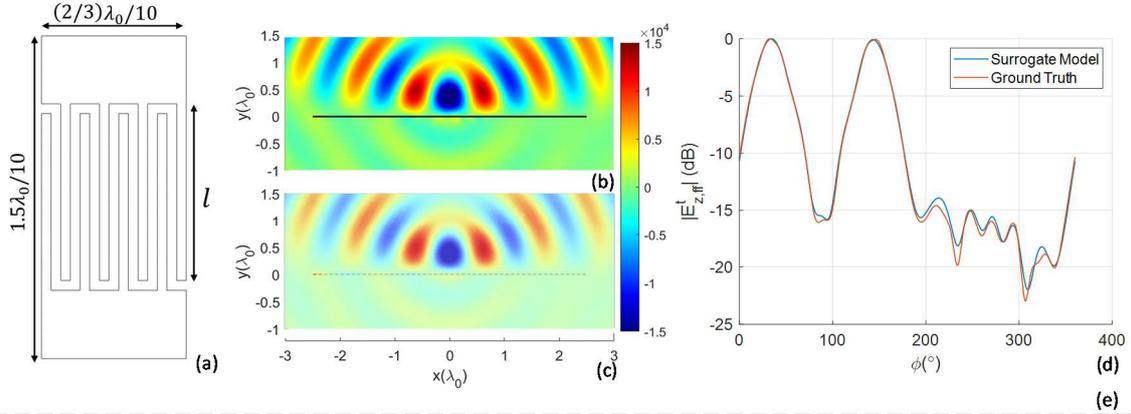

Fig. 1. (a) IDC geometry. (b) Near field of 50 element metasurface using new boundary condition. (c) Near field of 50 element metasurface from COMSOL Multiphysics. (d) Far field pattern comparison. (e) 50 element metasurface geometry.

### III. Aperiodic Metasurface Example

A metasurface consisting of 50 IDC's suspended in free-space and placed co-linearly along the $x$-axis was analyzed using the developed surrogate model and (3) expanded to the case of 50 meta-atoms. The metasurface is $5\lambda_0$ wide at the excitation frequency of 10GHz, hence, each unit cell is $\lambda_0/10$. The $l's$ of the IDC's are chosen to vary linearly from the minimum value of 0.0002mm on the left end to 0.004mm at the right end. The metasurface is excited by an infinite line source placed $\lambda_0/2$ above the center of the metasurface. The near and far total electric fields were calculated from the solution of (3) resulting in Figs. 1b and 1d.

Next, the metasurface was simulated in COMSOL using the actual IDC's (see Fig. 1e). The 50 IDC's were placed within a parallel plate waveguide and excited by the same line source. If the surrogate model represents the physics of a single unit cell accurately, then the COMSOL simulation should produce the same near and far fields as were previously obtained in Figs. 1b and 1d. The COMSOL results of the near field is shown in Fig. 1c and the far field is shown superimposed in Fig. 1d. The results are excellent, thus validating the new boundary condition.

### IV. Conclusion

A new homogenization-free boundary condition based on surrogate models was introduced for metasurface modelling. The boundary condition allows for the direct synthesis of the actual realized metasurface rather than of an approximate model thereby avoiding the problem of attempting to recreate the performance of the approximate model post-design, a step which is not always feasible. An example of a 50 IDC metasurface simulated using the new surrogate-based boundary condition was presented with excellent agreement with full-wave simulations thereby validating the approach.